\documentclass[amsthm]{PoS}

\usepackage{bm}
\usepackage{graphicx}
\usepackage{amsmath}
\usepackage{amsthm}
\usepackage{amsfonts}
\usepackage{amssymb}
\newcommand{\tht}{\textheight}
\newcommand{\ig}{\includegraphics}

\newcommand{\ba}{\overline{a}}
\newcommand{\bb}{\overline{b}}
\newcommand{\bc}{\overline{c}}

\newcommand{\bmu}{\overline{\mu}}
\newcommand{\bnu}{\overline{\nu}}
\newcommand{\btau}{\overline{\tau}}

\newcommand{\bpmu}{\overline{p}}
\newcommand{\bpnu}{\overline{p}}
\newcommand{\bptau}{\overline{p}}

\title{Group Theoretical Construction of Nucleon Operators using All-to-All Quark  Propagators}

\ShortTitle{All-to-all nucleon propagators}

\author{Lattice Hadron Physics Collaboration (LHPC): 
  R.G.~Edwards,$^a$ G.T.~Fleming,$^b$ 
  B.~Jo\'o,$^a$ \speaker{K.J.~Juge},$^c$ A.~Lichtl,$^d$ C.J.~Morningstar,$^e$ 
  D.G.~Richards,$^a$ 
  S.J.~Wallace $^f$\\
\llap{$^a$}Thomas Jefferson National Accelerator Facility, Newport News, VA 23606, USA\\
\llap{$^b$}Yale University, New Haven, CT 06520, USA\\
\llap{$^c$}Department of Physics, University of the Pacific, Stockton, CA 95211, USA\\
\llap{$^d$}RIKEN-BNL Research Center, Brookhaven National Laboratory, Upton, NY 11973 USA\\
\llap{$^e$}Department of Physics, Carnegie Mellon University, Pittsburgh, PA 15213, USA\\
\llap{$^f$}Department of Physics, University of Maryland, College Park, MD 20742, USA}

\abstract{We describe a method to construct irreducible baryon operators using all-to-all quark propagators. It was demonstrated earlier that a large basis of extended baryon operators on anisotropic, quenched lattices can be used to reliably extract the masses of 5 or more excited states in the nucleon channel. All-to-all quark propagators are expected to be needed when studying these excited states on light, dynamical configurations because contributions from multi-particle states are expected to be significant. The dilution method is used to approximate the all-to-all quark propagators. Low-lying eigenmodes can also be used if necessary. For efficient computation of matrix elements of the interpolating operators, the algorithms should exploit the fact that many extended baryon operators can be obtained from the different linear combinations of three-quark colour-singlet operators. The sparseness of the diluted noise vectors also afford several computation simplifications. Some preliminary results are presented for nucleon effective masses.}

\FullConference{The XXV International Symposium on Lattice Field Theory\\
		            July 30-4 August 2007\\
		            Regensburg, Germany}

\begin{document}

\section{Introduction}

The theoretical determination of the hadron spectrum from first principles is one of the goals of the Lattice Hadron Physics Collaboration. Lattice QCD calculations have provided reasonable agreement with the physical ground state masses of different baryons (for a review, see Ref.~\cite{Ishikawa:2004nm}) and the excited state baryon spectra in the quenched approximation has been reported by several groups ~\cite{Zhou:2006xe,Burch:2006cc,Sasaki:2005uq,Sasaki:2005ug,Basak:2004hr,Zanotti:2003fx}. In the presence of light, dynamical quarks, it is expected that multi-particle states will have to be included for a thorough study of the excited state spectra of hadrons.

One of the difficulties of studying multi-particle states is the formidable number of contractions that one has with conventional point-to-all quark propagators. All-to-all quark propagators solve this problem since colour and spin contractions are performed on the sinks and sources separately. One also gains access to disconnected diagrams which are difficult to evaluate with conventional propagators. In this work, we test the ``dilution" method of Ref.~\cite{Foley:2005ac} for approximating the all-to-all quark propagators.

The low-lying nucleon spectrum obtained through group-theoretical projections (\cite{Basak:2005aq}) in the quenched approximation has been reported in Refs.~\cite{Lichtl:2006dt,Juge:2006gr}. Our aim is to compare the point-to-all and all-to-all constructions of these nucleon correlation matrices. We find comparable (or better) signals in the single-particle sector and verify that diluted all-to-all propagators can be used to study excited baryons.

\section{Construction of Operators/Correlators}

\subsection{Point-to-All Construction}

We follow the group-theoretical construction of baryon operators outlined in Ref.~\cite{Basak:2005aq}. A nucleon correlation matrix is formed by combining various colour contracted three-quark propagators, $\widetilde{G}^{(uud)(p\overline{p})}_{(\mu|\bar{\mu})(\nu|\bar{\nu})(\tau|\bar{\tau})}$ where the indices denote the spin and displacements of the three quarks (sink|source notation),
\begin{eqnarray}\nonumber
 C^{(N)}_{ij}(t) &=& c^{(i)}_{\mu\nu\tau}
\, \bc^{(j)}_{\bmu\bnu\btau}
\Bigl\{
\widetilde{G}^{(uud)(p\overline{p})}_{ (\mu  \vert \bmu)
(\nu  \vert \bnu)
(\tau \vert \btau)}
+\widetilde{G}^{(uud)(p\overline{p})}_{ (\mu  \vert \bnu)
(\nu  \vert \bmu)
(\tau \vert \btau)} 
-\widetilde{G}^{(uud)(p\overline{p})}_{ (\mu  \vert \btau)
(\nu  \vert \bnu)
(\tau \vert \bmu)}
-\widetilde{G}^{(uud)(p\overline{p})}_{ (\mu  \vert \bnu)
 (\nu  \vert \btau)
 (\tau \vert \bmu)} \nonumber\\
&-&\widetilde{G}^{(uud)(p\overline{p})}_{ (\nu  \vert \bnu) 
 (\tau \vert \bmu)
 (\mu  \vert \btau)}
-\widetilde{G}^{(uud)(p\overline{p})}_{ (\nu  \vert \bmu)
 (\tau \vert \bnu) 
 (\mu  \vert \btau)}
+\widetilde{G}^{(uud)(p\overline{p})}_{ (\tau \vert \btau)
 (\nu  \vert \bnu)
 (\mu  \vert \bmu)}
+\widetilde{G}^{(uud)(p\overline{p})}_{ (\tau \vert \bnu)
 (\nu  \vert \btau)
 (\mu  \vert \bmu)} 
\Bigr\}
\end{eqnarray}
with the various projection coefficients, $c^{(i)}_{\mu\nu\tau}$ and 
$\bc^{(j)}_{\bmu\bnu\btau}$. Our notation follows closely that of Ref.~\cite{Basak:2005aq}. The (smeared) three-quark propagators are formed from colour contracting the source and sink colour indices of the quark propagators,
%\begin{eqnarray}\nonumber
\begin{equation}
\widetilde{G}^{(ABC)(p\overline{p})}_{(\alpha i\vert\overline{\alpha}\overline{i})
(\beta j\vert\overline{\beta}\overline{j})
(\gamma k\vert\overline{\gamma}\overline{k})}(t)%\nonumber\\
=\sum_{\bm{x}}
\varepsilon_{abc}\,\varepsilon_{\ba\bb\bc}
\ \widetilde{Q}^{(A)}_{a\alpha i p\vert\ba
 \overline{\alpha}\overline{i}\overline{p}}(\bm{x},t\vert\bm{x}_0,0)%\nonumber\\
  \widetilde{Q}^{(B)}_{b\beta j p\vert\bb
   \overline{\beta}\overline{j}\overline{p}}(\bm{x},t\vert\bm{x}_0,0)
\ \widetilde{Q}^{(C)}_{c\gamma k p\vert\bc
   \overline{\gamma}\overline{k}\overline{p}}(\bm{x},t\vert\bm{x}_0,0)
\quad 
\end{equation}
%\end{eqnarray}
where
$
\widetilde{Q}^{(A)}_{a\alpha i p\vert\ba
 \overline{\alpha}\overline{i}\overline{p}}(\bm{x},t\vert\bm{x}_0,0)
$
are the (smeared) quark propagators as defined in Ref.~\cite{Basak:2005aq}. 

\subsection{All-to-All Construction}
An expression for the nucleon propagator in terms of all-to-all quark propagators is  obtained by replacing each quark propagator matrix, $\widetilde{Q}^{(A)}$, in the previous expression with 
$$
\widetilde{Q}_{a\alpha i p\vert\ba
\overline{\alpha}\overline{i}\overline{p}}(\bm{x},t\vert\bm{x}_0,t_0)=\lim_{N\rightarrow\infty}\frac{1}{N}\sum_{A=1}^N\sum_d^{N_{dil}}\widetilde{\phi}^{(d)}_{[A]a\alpha ip}(\bm{x},t)\otimes\widetilde{\eta}^{(d)\dagger}_{[A]\overline{a}\overline{\alpha}\overline{i}\overline{p}}(\bm{x}_0,t_0)
$$
where $\widetilde{\eta}^{(d)}_{[A]}$ are the (smeared) diluted random noise sources for source $A$ and $\widetilde{\phi}^{(d)}_{[A]}$ are the corresponding (smeared) solutions (see Ref.~\cite{Foley:2005ac}). The summation over the dilution set is no more than  $N_{dil}=N_tN_cN_\sigma$ in this study and we take $N=1$ (separately) for each of the three quark propagators. (In other words, each of the three quark propagators are estimated independently.) The nucleon propagator then takes the form (after some symmetry transformations),
\begin{eqnarray}\nonumber
 C^{(N)}_{IJ}(t,t_0) &=& \frac{1}{N^3}\sum_{A}^{N}\sum_{B}^{N}\sum_{C}^{N}\sum_{\tilde{i}}^{N^A_{dil}}\sum_{\tilde{j}}^{N^B_{dil}}\sum_{\tilde{k}}^{N^C_{dil}}\sum_{\bm{x}}\sum_{\bm{x}_0}
c^{(I)}_{\mu\nu\tau}B^{\tilde{i}\tilde{j}\tilde{k}}_{[ABC]\mu\nu\tau}(\bm{x},t)\\\nonumber
&&
\bc^{(J)}_{\bmu\bnu\btau}\Bigl\{\,
 2\overline{B}^{\tilde{i}\tilde{j}\tilde{k}}_{[ABC]\bmu\bnu\btau}
+2\overline{B}^{\tilde{k}\tilde{j}\tilde{i}}_{[CBA]\bmu\bnu\btau}%\\\nonumber
%&&\hspace*{32mm}
-\overline{B}^{\tilde{j}\tilde{i}\tilde{k}}_{[BAC]\bmu\bnu\btau}
-\overline{B}^{\tilde{i}\tilde{k}\tilde{j}}_{[ACB]\bmu\bnu\btau}
-\overline{B}^{\tilde{k}\tilde{i}\tilde{j}}_{[CAB]\bmu\bnu\btau}
-\overline{B}^{\tilde{j}\tilde{k}\tilde{i}}_{[BCA]\bmu\bnu\btau}\,\Bigr\}(\bm{x}_0,t_0)\\\nonumber
&\equiv&\frac{1}{N^3}\sum_{A}^{N}\sum_{B}^{N}\sum_{C}^{N}\sum_{\tilde{i}}^{N^A_{dil}}\sum_{\tilde{j}}^{N^B_{dil}}\sum_{\tilde{k}}^{N^C_{dil}}{\mathcal O}_I(t;\tilde{i}\tilde{j}\tilde{k}){\overline{\mathcal O}}_J(t_0;\tilde{i}\tilde{j}\tilde{k})
\end{eqnarray}
where
\begin{eqnarray}\nonumber
B^{\tilde{i}\tilde{j}\tilde{k}}_{[ABC]\mu\nu\tau}(\bm{x},t)&\equiv&\varepsilon_{abc}\bigl(\widetilde{D}^{(p)}_{i_\mu}\ \widetilde{\phi}^{(\tilde{i})}_{[A]\mu}\bigr)_a(\bm{x},t)\bigl(\widetilde{D}^{(p)}_{i_\nu}\ \widetilde{\phi}^{(\tilde{j})}_{[B]\nu}\bigr)_b(\bm{x},t)\bigl(\widetilde{D}^{(p)}_{i_\tau}\ \widetilde{\phi}^{(\tilde{k})}_{[C]\tau}\bigr)_c(\bm{x},t)\\\nonumber
\overline{B}^{\tilde{i}\tilde{j}\tilde{k}}_{[ABC]\bmu\bnu\btau}(\bm{x}_0,t_0)&\equiv&\varepsilon_{\ba\bb\bc}\bigr(\widetilde{D}^{(\bpmu)}_{i_{\bar{\mu}}}\widetilde{\eta}^{(\tilde{i}) }_{[A]\bmu}\bigr)^{\dagger}_{\ba}(\bm{x}_0,t_0)\bigr(\widetilde{D}^{(\bpnu)}_{j_{\bar{\nu}}}\widetilde{\eta}^{(\tilde{j}) }_{[B]\bnu}\bigr)^{\dagger}_{\bb}(\bm{x}_0,t_0)\bigr(\widetilde{D}^{(\bptau)}_{k_{\bar{\tau}}}\widetilde{\eta}^{(\tilde{k}) }_{[C]\btau}\bigr)^{\dagger}_{\bc}(\bm{x}_0,t_0)
\end{eqnarray}
are colour-singlet three-quark operators and
\begin{equation}\nonumber
 \widetilde{D}_j^{(p)}(x,x^\prime) =
 \widetilde{U}_j(x)\ \widetilde{U}_j(x\!+\!\hat{j})\dots 
   \widetilde{U}_j(x\!+\!(p\!-\!1)\hat{j})\delta_{x^\prime,x+p\hat{j}}.
\end{equation}
The tilde's indicate that the fields have been smeared.

The task of computing correlation functions has therefore been reduced to computing the source and sink operators, ${\overline{\mathcal O}}_J(t_0;\tilde{i}\tilde{j}\tilde{k})$ and ${\mathcal O}_I(t;\tilde{i}\tilde{j}\tilde{k})$ separately and then contracting the remaining dilution indices to form correlation functions. The latter step does not require large resources and hence can be performed as a post-production operation. This scenario is very different from the point-to-all construction which requires one to make the correlation functions directly from the quark propagators.

\subsection{Optimization}
The dominant computational effort is to make the colour-singlet three-quark  operators, $B^{\tilde{i}\tilde{j}\tilde{k}}_{[ABC]\mu\nu\tau}$, for each dilution index combination. We are interested in making a correlation matrix with many baryon operators. There are three-quark operators that are present in a large number of different baryon operators. This is especially the case if we were to make the baryon operators for all different isospins, irreps and rows of irreps at once. Therefore, it is crucial that we only compute each three-quark operator once in order to optimize the computational time. In this first study, we shall only consider common the three-quark operators that have the same spin indices. Common quark displacements will be considered in a subsequent study.

\section{Simulation/Results}

\subsection{Parameters}
We have used the same configurations as in our earlier studies with the conventional, point-to-all quark propagators (Ref.~\cite{Lichtl:2006dt,Juge:2006gr}) for comparison purposes. These are the anisotropic ($\frac{a_s}{a_t}~=~3$) quenched, Wilson gauge action ($\beta=6.1$) on $12^3\times48$ lattices. Anisotropic Wilson fermions with pion masses of roughly $700\ \text{MeV}$ were used for the valence quarks.

\subsection{Operators}
In this first test, we have focused on the nucleon $G_{1g}$ channel with one operator of each ``displacement type" (apart from the single-site case where we have tried all three). These are the  Single-Site (SS-0,1,2), Singly-Displaced (SD-0), Doubly-Displaced-I (DDI-0), Doubly-Displaced-L (DDL-1) and Triply-Displaced-T (TDT-3) type operators. We refer to Ref.~\cite{Basak:2005aq} for further details on these operators.

\subsection{Cost Comparison}
Twenty configurations were analyzed for time-dilution and time-spin-dilution schemes. Time-spin-colour-dilution has been tested on a single configuration thus far. Since the traditional method requires one to compute the correlation functions directly, we compare the number of quark inversions required for each computation as an indication of the cost of the simulation time. We note, however, that the construction of operators (correlation functions) consumes most of the simulation time when using all-to-all (point-to-all) propagators for the heavy quark mass used here.

%\subsubsection{Time-dilution}
\indent{\bf Time-dilution}
The lattice size in the time-direction of the anisotropic lattices used here was $N_t=48$. The total number of inversions is then $144=N_t\times3$ (for three quarks) to generate the time-diluted all-to-all quark propagator set. In the conventional point-to-all construction, one needs $N_c\times N_\sigma\times7=84$ inversions in order to make all of the displaced operators.

%\subsubsection{Time-Spin-dilution}
\indent{\bf Time-Spin-dilution}
The number of inversions in this scheme is increased by a factor of 4 to $576=3\times N_\sigma\times N_t\simeq7\times84$, when we dilute in spin as well as in time. The cost of extra inversions for the time-spin-diluted all-to-all propagators is then nearly a factor of 7 compared to the point-to-all construction. 

\subsection{Effective Masses}
The effective masses are collectively shown in Fig.~1 for time-dilution and Fig.~2 for  time-spin-dilution. The effective masses with time-dilution alone have large error bars indicating that a higher dilution scheme is necessary. 
%. 
On the other hand, the time-spin-dilution scheme shows error bars which are already compatible with the point-to-all effective masses on the same configurations. It is important to note that even though the cost of generating extra quark propagators is not compensated by a reduction in the error bars, the main purpose of using  all-to-all propagators is to construct explicit multi-particle operators and correlation-functions on dynamical configurations, not just to reproduce the quenched spectra of Ref.~\cite{Lichtl:2006dt,Juge:2006gr}. It is sufficient to have error bars which are comparable to those of the point-to-all method for the purpose of analyzing single particle states, although further reduction can be achieved by going to a higher dilution scheme if necessary (see Fig.~3 for an example of colour dilution). We also note that the added statistics obtained by having source operators on every timeslice on the lattice have been utilized in the all-to-all simulations (see Fig.~4). 

\subsection{Diagonalization/Excited States}
The main idea behind using a matrix of correlators with many different operators is to find a good basis of operators which overlap well with excited states. Since the diagonalization procedure involves correlation matrices on two different timeslices, there is a possiblity that the stochastic noise on the two timeslices will make the diagonalization unstable. So it is necessary to test whether time-diluted noise vectors can be used in the variational method. We have taken one operator of each kind (just the first operator for SS) and diagonalized the correlation matrix on timeslice 5. The effective masses of the first three states are shown in Fig.~5. Figure~6 shows the corresponding results for the point-to-all method. One can see that the diagonalization procedure based on the all-to-all method is as stable as the one based on the point-to-all method. The correlation matrix for time-dilution alone was too noisy to be diagonalized. This is an example where time-dilution alone is not sufficient to even reproduce the point-to-all results.

%%%%%%%%%%%%%%
\begin{figure}[t]
\begin{center}
\begin{tabular}{cc}
\begin{minipage}{73mm}
\begin{center}
\ig[height=0.3\tht]{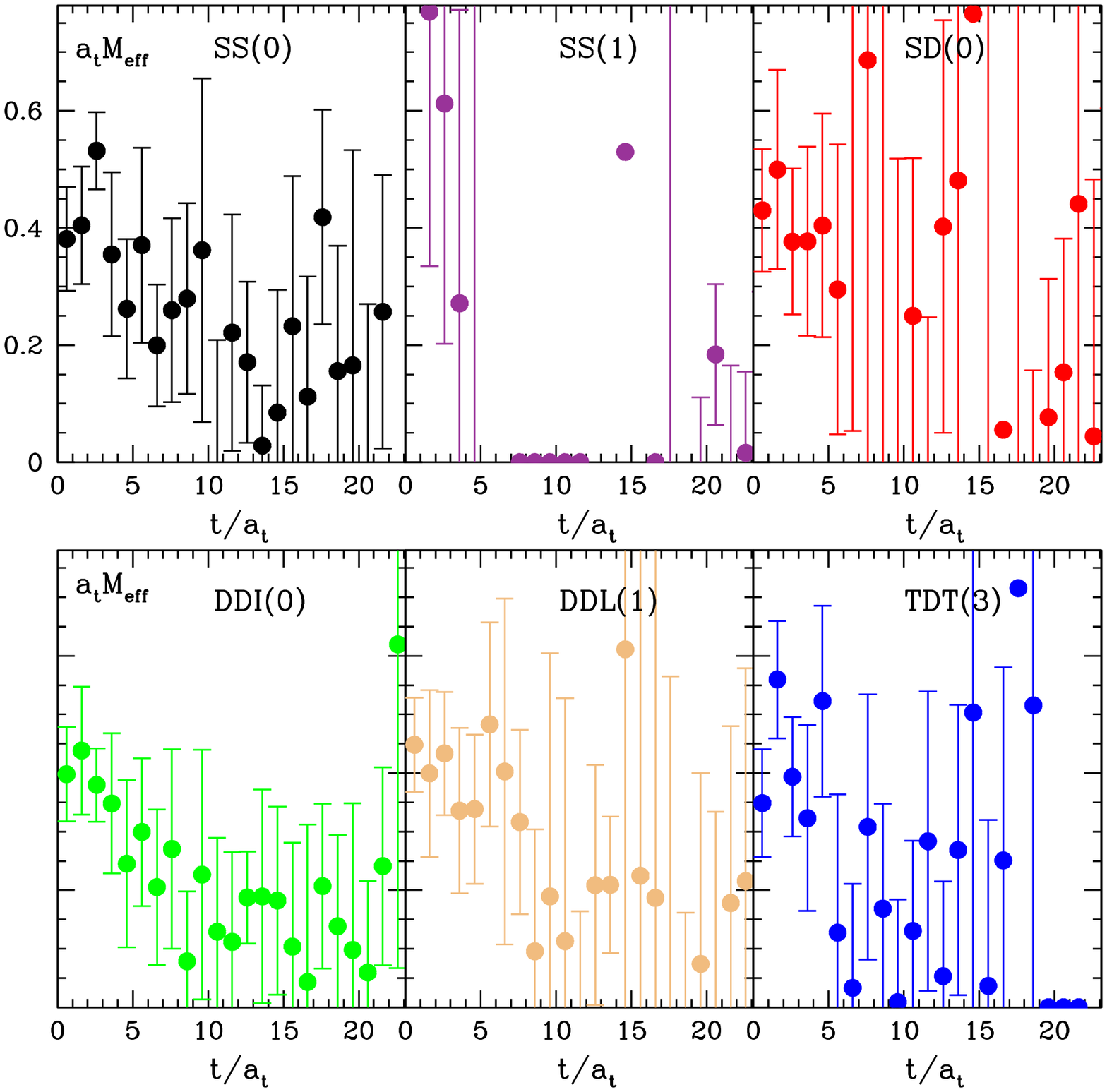} 
\caption{Time-diluted effective masses for various operators in the $G_{1g}$ irrep with isospin 1/2.}
\end{center}
\end{minipage}&
\label{fig:Tf}
\begin{minipage}{73mm}
\begin{center}
\ig[height=0.3\tht]{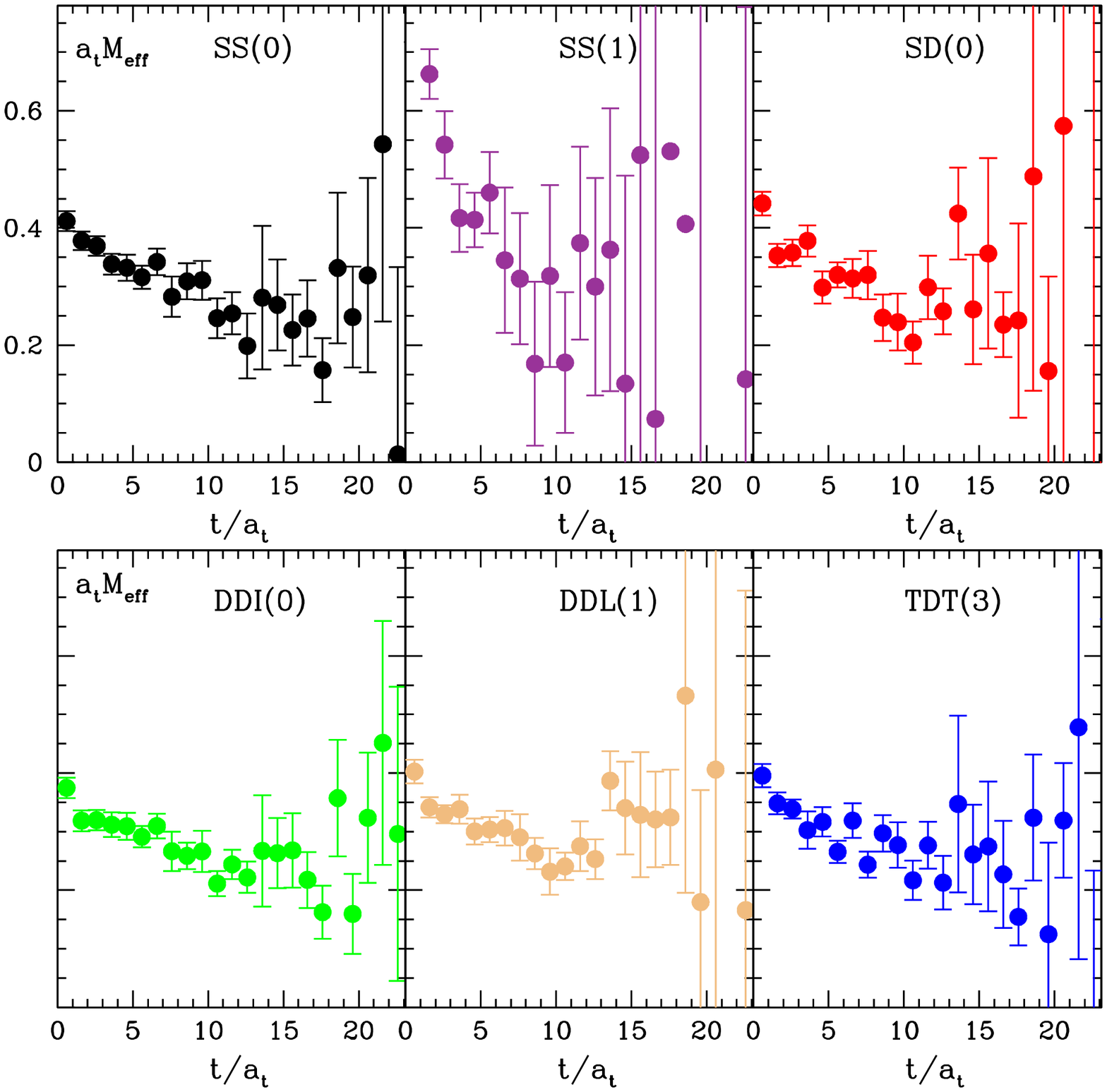}
\caption{Time-Spin-diluted effective masses for various operators in the $G_{1g}$ irrep with isospin 1/2.}
\label{fig:TfGf}
\end{center}
\end{minipage}
\end{tabular}
\end{center}
\end{figure}
%%%%%%%%%%%%%%

%%%%%%%%%%%%%%
\begin{figure}[t]
\begin{center}
\begin{tabular}{cc}
\begin{minipage}{73mm}
\begin{center}
\ig[height=0.3\tht]{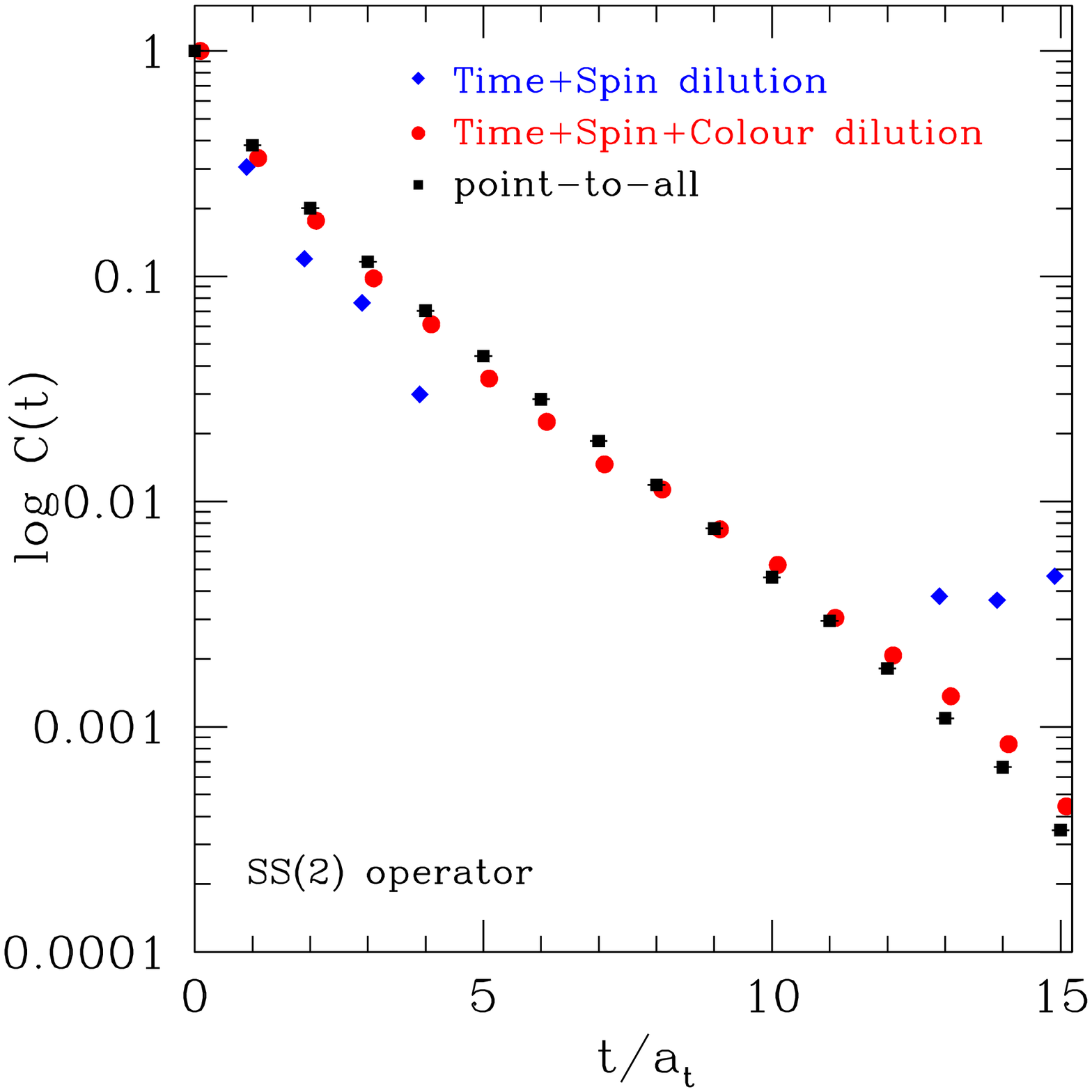}
\label{fig:TfSfCf}
\caption{The nucleon propagator for time-spin-dilution, time-spin-colour-dilution and point-to-all quark propagators for the Single-Site-\#2 operator in the $G_{1g}$ irrep with isospin 1/2 on a single configuration.} 
\end{center}
\end{minipage}&
\label{fig:Tf}
\begin{minipage}{73mm}
\begin{center}
\ig[height=0.3\tht]{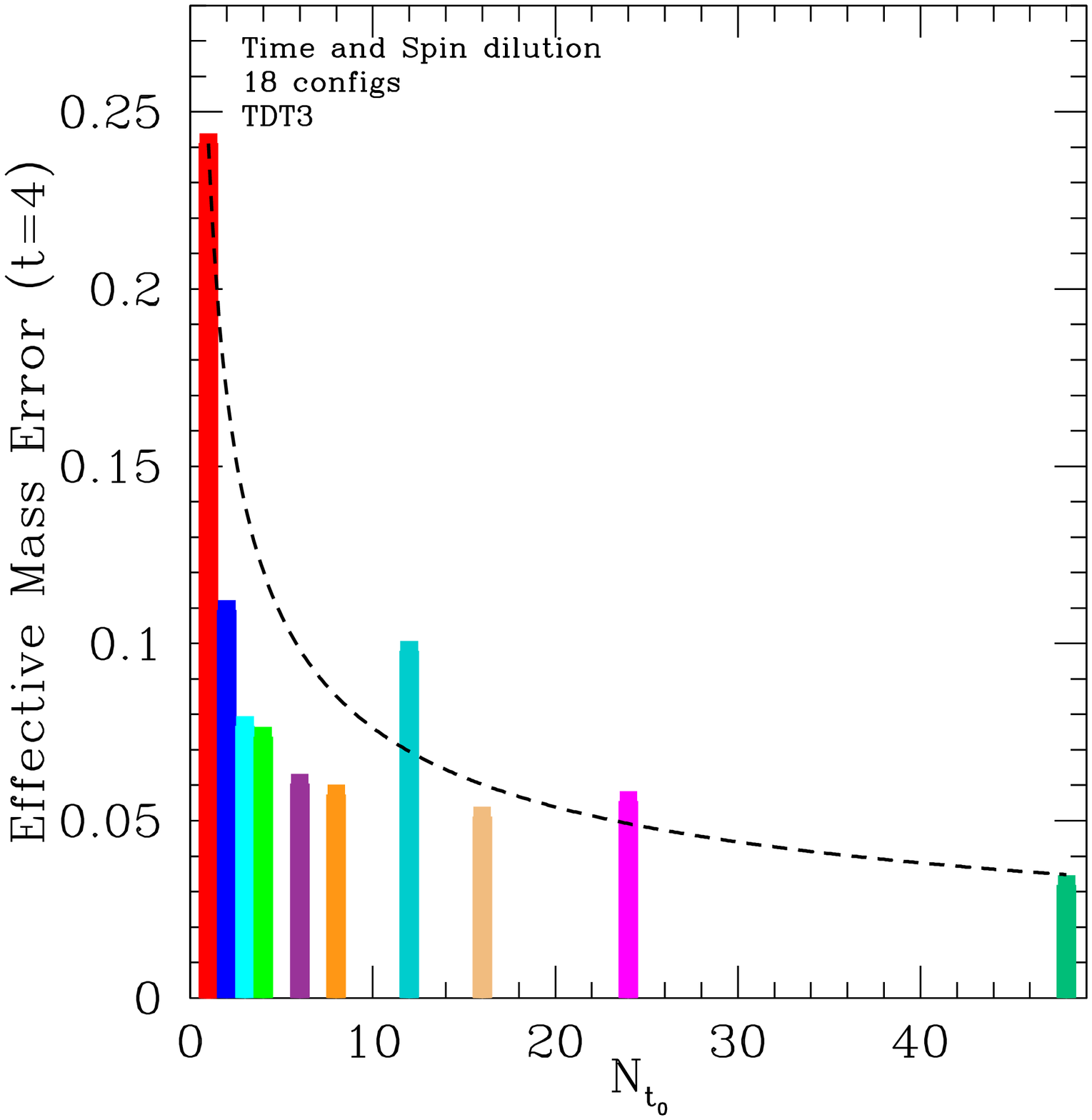}
\caption{The error of the nucleon effective mass for the SS(0) operator on timeslice 4 as a function of the number of sources used on each configuration. The dashed line is $\frac{1}{\sqrt N_{t_0}}$.} 
\label{fig:Nt0}
\end{center}
\end{minipage}
\end{tabular}
\end{center}
\end{figure}
%%%%%%%%%%%%%%

%%%%%%%%%%%%%%
\begin{figure}[t]
\begin{center}
\begin{tabular}{cc}
\begin{minipage}{73mm}
\begin{center}
\ig[height=0.3\tht]{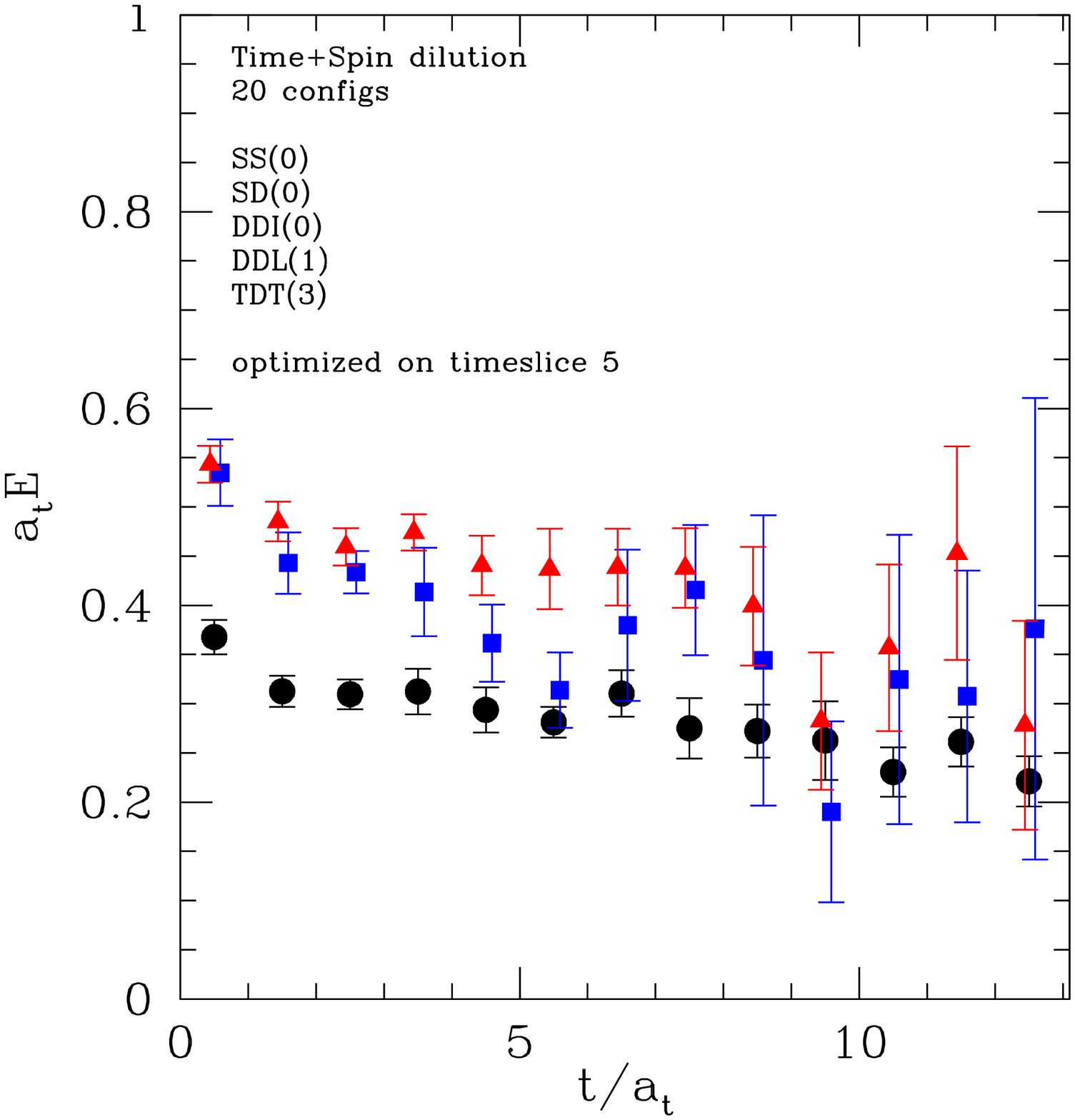} 
\caption{Effective masses of the optimized (diagonal) correlators using time and spin diluted all-to-all quark propagators.}
\label{fig:TfGf}
\end{center}
\end{minipage}&
\begin{minipage}{73mm}
\begin{center}
\ig[height=0.3\tht]{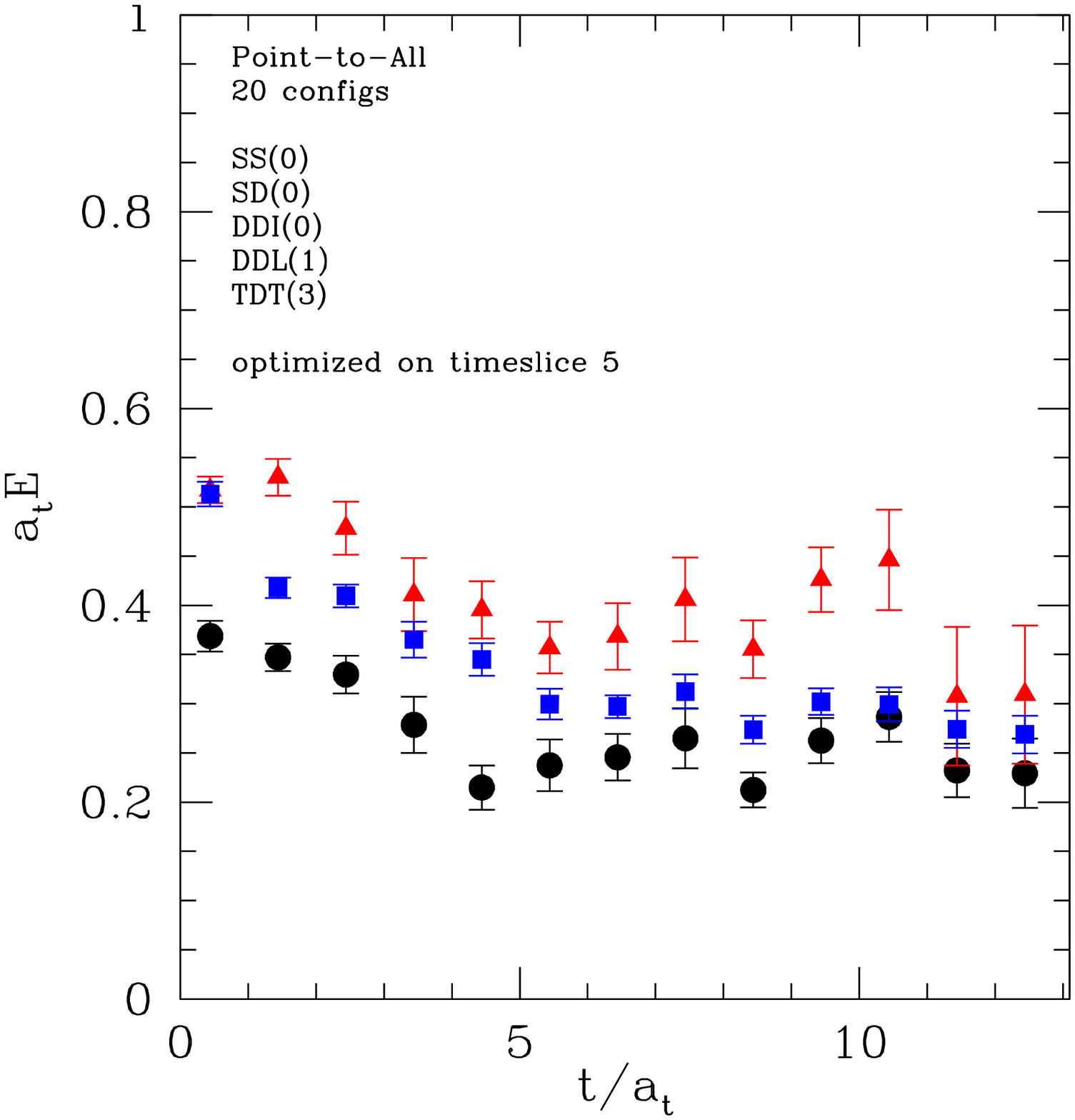}
\caption{Effective masses of the optimized (diagonal)\ correlators using conventional point-to-all quark propagators.}
\label{fig:point}
\end{center}
\end{minipage}
\end{tabular}
\end{center}
\end{figure}
%%%%%%%%%%%%%%

\section{Summary}
The efficacy of nucleon operator construction using group-theoretical projections and noise-diluted all-to-all quark propagators has been illustrated for selected operators. It was found that time and spin dilutions were both necessary to obtain signals for the nucleons that are comparable to those of the point-to-all method. We also have a good indication that diluting in colour will further reduce the stochastic noise for some of the noisier operators.

The independent $Z_4$ random sources on each timeslice coming from diluting the stochastic source vector did not pose a problem for extracting excited states with time-spin dilution. The signal for the first three states was clearly seen with the limited number of operators and configurations which were used in this first study. The importance of studying the dilution scheme was illuminated by the failure of only diluting in time.

The next step in our study is to construct explicit multi-particle operators with diluted, all-to-all propagators. This will enlarge the basis of operators which may be used to obtain a good overlap with excited states on dynamical configurations with ``light" quark masses. It may also become useful to include low-lying eigenmodes together with dilutions (the hybrid method) when studying lighter quarks. Work is in progress in these directions.

\section*{Acknowledgements}
K.J.J. would like to thank Mike Peardon for many discussions on using diluted, all-to-all quark propagators for baryon spectroscopy.
This work has been partially supported by National Science Foundation awards PHY-0653315/PHY-0704171 and by the Department of Energy under contracts DE-AC05-06OR23177 and DE-FG02-93ER-40762.

%\bibliography{pacific}

\end{document}